\documentclass[aps, prb, amsmath, twocolumn, longbibliography, 10pt]{revtex4-1}

\usepackage{graphicx} 
\usepackage{multibib} 
\usepackage{xcolor}
\newcites{supp}{ } 

\begin{document}

\title{High-Bandwidth Extended-SWIR GeSn Photodetectors on Silicon Achieving Ultrafast Broadband Spectroscopic Response}

\author{M. R. M. Atalla}
\affiliation{Department of Engineering Physics, \'Ecole Polytechnique de Montr\'eal, C.P. 6079, Succ. Centre-Ville, Montr\'eal, Qu\'ebec, Canada H3C 3A7}
\author{S. Assali}
\affiliation{Department of Engineering Physics, \'Ecole Polytechnique de Montr\'eal, C.P. 6079, Succ. Centre-Ville, Montr\'eal, Qu\'ebec, Canada H3C 3A7}
\author{S. Koelling}
\affiliation{Department of Engineering Physics, \'Ecole Polytechnique de Montr\'eal, C.P. 6079, Succ. Centre-Ville, Montr\'eal, Qu\'ebec, Canada H3C 3A7}
\author{A. Attiaoui}
\affiliation{Department of Engineering Physics, \'Ecole Polytechnique de Montr\'eal, C.P. 6079, Succ. Centre-Ville, Montr\'eal, Qu\'ebec, Canada H3C 3A7}
\author{O. Moutanabbir}
\email{oussama.moutanabbir@polymtl.ca}
\affiliation{Department of Engineering Physics, \'Ecole Polytechnique de Montr\'eal, C.P. 6079, Succ. Centre-Ville, Montr\'eal, Qu\'ebec, Canada H3C 3A7}

\begin{abstract}
The availability of high-frequency pulsed emitters in the $2 \text{--} 2.5\,\mu$m wavelength range paved the way for a wealth of new applications in ultrafast spectroscopy, free-space and fiber-optical communications, surveillance and recognition, artificial intelligence, and medical imaging. However, developing these emerging technologies and their large-scale use depend on the availability of high-speed, low-noise, and cost-effective photodetectors. With this perspective, here we demonstrate GeSn photodiodes grown on silicon wafers featuring a high broadband operation covering the extended-SWIR range with a peak responsivity of 0.3 A/W at room temperature. These GeSn devices exhibit a high bandwidth reaching 7.5 GHz at 5 V bias with a 2.6 $\mu$m cutoff wavelength, and their integration in ultrafast time-resolved spectroscopy applications is demonstrated. In addition to enabling time-resolved electro-luminescence at 2.3 $\mu$m, the high-speed operation of GeSn detectors was also exploited in the diagnostics of ultra-short pulses of a supercontinuum laser with a temporal resolution in the picosecond range at 2.5 $\mu$m. Establishing these capabilities highlights the potential of manufacturable GeSn photodiodes for silicon-integrated high-speed extended-SWIR applications. 
\end{abstract}

\maketitle

Developing high-speed photodetectors (PDs) operating at room-temperature in the extended-SWIR (e-SWIR) spectral range is critical to implement a variety of applications including high-resolution active light detection and ranging (LIDAR), time-resolved spectroscopy, environmental monitoring of greenhouse gases, medical optical tomography, and new generation optical communication systems.\cite{wun2016,Williams2017,Kim2021} State-of-the-art photodetectors in this wavelength range with a bandwidth above 5 GHz consist exclusively of In-rich InGaAs on InP,\cite{Ye2015,joshi2008high,yang2013} InGaAs/GaAsSb on InP,\cite{chen2018,tossoun2018,chen2019hi} and GaInAsSb on GaSb.\cite{andreev2013high,wun2016} These devices face cost and scalability challenges and exhibit low bandwidth above 2.3 \text{$\mu$}m that does not exceed 6 GHz.\cite{wun2016} Establishing silicon-integrated high-speed detectors is an attractive paradigm for large-scale fabrication, cost effectiveness, and compatibility with complementary metal-oxide semiconductor (CMOS) processing.\cite{geis2007cmos} In this regard, recent attempts were made using ion-implanted silicon to engineer defect-mediated absorption yielding PDs with a cutoff wavelength of 2.3 \text{$\mu$}m and a 15 GHz bandwidth at 2 \text{$\mu$}m.\cite{grote201210,souhan2014si+} However, these defect-based PDs are limited by their low responsivity, large device footprint, and high operation bias.\cite{ackert2015high} 

\medskip

GeSn semiconductors epitaxially grown on silicon substrates provide a serious alternative to implement monolithic, silicon-integrated devices, where the bandgap directness and energy can be tuned by controlling tin content.\cite{soref2010mid,moutanabbir2021mono} As a matter of fact, GeSn has been recently integrated in the fabrication of detectors\cite{soref2015enabling,xu2019high,tran2019study,talamas2021} and emitters\cite{elbaz2020ultra,zhou2020elec,chtien2019} in the 2\text{--}2.5 \text{$\mu$}m range and beyond. For instance, recent Ge/GeSn/Ge PIN PD results showed a high potential by demonstrating a 2.5 \text{$\mu$}m cutoff and a 1.78 GHz bandwidth at 2 \text{$\mu$}m.\cite{tran2019study} Nevertheless, this bandwidth remains more than 3 times less than the bandwidth of III-V PDs at similar cutoff wavelength. Herein, this work demonstrates all-GeSn PIN heterostructures grown on silicon substrates yielding higher speed, free-space PDs. These devices exhibit a broadband spectral responsivity and a high bandwidth of 7.5 GHz at a bias of 5 V, a 2.6 \text{$\mu$}m cutoff wavelength, and a 0.3 A/W responsivity at room temperature. The integration of these devices in ultrafast spectroscopy is also described. The high-speed operation of GeSn PDs was exploited to demonstrate ultra-short pulse characterization revealing pulse duration, intensity, and spectral distribution of a pulsed supercontinuum laser using a Fourier Transform InfraRed (FTIR) spectrometer reaching a hitherto unmatched temporal resolution in the picosecond range at $2.5\,\mu$m. These PDs were also integrated in time-resolved electro-luminescence spectroscopy to determine the carrier lifetime of a $2.3\,\mu$m light emitting diode (LED).

\begin{figure*}[t]
    \centering
    \includegraphics[scale=0.77]{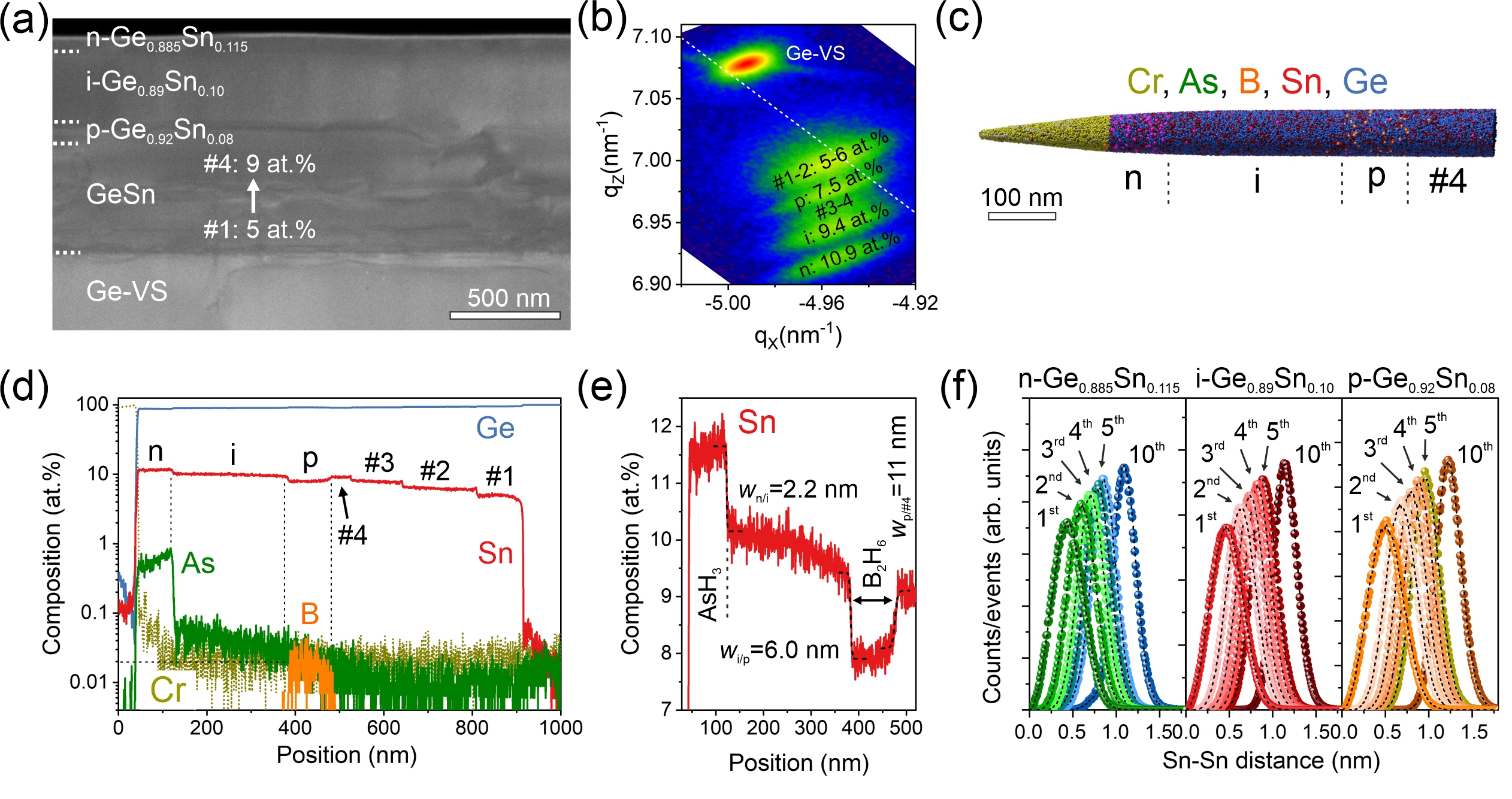}
    \caption{{\bf Material growth and characterization.} (a) TEM image showing the CVD grown PIN stack on top of GeSn and Ge-VS buffer layers. (b) RSM map depicting the strain and Sn at.$\%$ in different layers. (c) 3-D atom-by-atom reconstruction of the entire atom probe tip indicating the elemental distribution of the As, B, Ge, Sn and Cr atoms. (d) Concentration profiles of different elements across the GeSn stack showing the width of the i-layer to be $300\,\text{nm}$. (e) Sn profile depicting the width of the PIN interfaces. (f) Nearest-neighbor statistical analysis of the PIN layers confirming the perfect random GeSn alloy and the absence of Sn clustering.} 
\end{figure*}

\medskip

\noindent {\bf Results and discussion}

\noindent {\bf Growth and characterization of GeSn epilayers}


GeSn PIN heterostructures were grown on a 4-inch Si ($100$) wafers using a Ge virtual substrate (Ge-VS) in a low-pressure chemical vapor deposition (CVD) reactor (Methods).  Fig. 1(a) displays a cross-sectional transmission electron micrograph (TEM) of a heterostructure at a total thickness of ~960 nm. Dislocations are mainly observed in the lower portion of the stacking that extends for a thickness of ~600 nm to reach the p-GeSn layer, while the i-GeSn and n-GeSn regions are of high crystallinity. The gliding of misfit dislocations at the interface between GeSn layers having different Sn content is promoted rather than the nucleation and propagation of threading dislocations through the stacking.\cite{aubin2017} X-ray diffraction (XRD) Reciprocal Space Mapping (RSM) measurements around the asymmetrical (224) reflection were carried out to evaluate the composition and strain in the grown GeSn layers (Fig. 1(b)). The observed small in-plane tensile strain below $0.2\,\%$ in Ge-VS results from the thermal cyclic annealing, while the lattice-mismatched growth of GeSn leads to a compressive strain in the epilayers. Plastic relaxation during growth, as evidenced by dislocation gliding in Fig. 1(a), yields a very low residual compressive strain of about $-\,0.2\,\%$ in the buffer layers \#1-3. The strain increases slightly to  $\sim-\,0.3\,\%$ in layer \#4 and in p-Ge$_{0.924}$Sn$_{0.076}$. As Sn content increases, a higher strain of  $\sim\,-\,0.5\,\%$ and $-\,0.7\,\%$ is recorded in the i-Ge$_{0.909}$Sn$_{0.091}$ and n-Ge$_{0.884}$Sn$_{0.106}$ layers, respectively.

\medskip

Before any device processing, the GeSn material quality and structural properties down to the atomic-level were investigated further using atom probe tomography (APT). The three-dimensional (3D) reconstruction of the GeSn PIN layers is shown in Fig. 1(c), while the extracted compositional profiles are displayed in Fig. 1(d). Note that chromium (Cr) was deposited on as-grown samples to facilitate the preparation of the APT tips.\cite{koelling2020} Although the small difference in atomic mass between As and Ge atoms makes decoupling As from Ge signal in APT challenging, a clear As profile is visible in the top GeSn layer with a background signal across the heterostructure. The interfacial width (w) in the PIN layers is evaluated by fitting the compositional profiles (Methods). We first focus on the Sn profile, where uniform compositions of $5$ at.$\%$ (\#1), $6$ at.\% (\#2), $8$ at.\% (\#3), and $9$ at.\% (\#4) are obtained in the buffer layers. The Sn incorporation is controlled by a $10\,^\circ\text{C}$-step temperature reduction during growth, resulting in sharp interfaces with w = $1\text{--}3$ nm. In the p-GeSn layer, the Sn content decreases from $9$ to $8$ at.\%, but with a much broader interface (w$_{(p/\#4)}\sim$11 nm) that results from the slow incorporation of B in the lattice. A sharper transition at the i-Ge$_{0.90}$Sn$_{0.10}$/p-Ge$_{0.924}$Sn$_{0.076}$ interface (w$_{(i/p)}\sim6.0$ nm) is observed along with a uniform composition. A relatively abrupt n-Ge$_{0.884}$Sn$_{0.106}$/i-Ge$_{0.90}$Sn$_{0.10}$ interface is recorded for both Sn (w$_{(\text{Sn}-n/i)}\sim2.2$ nm) and As profiles (w$_{(\text{As}-n/i)}\sim4.1$ nm). Moreover, the Sn content increases by $1$ at.\% in n-Ge$_{0.885}$Sn$_{0.115}$ (Fig. 1(e)), which stands in sharp contrast to earlier observations, where As doping was reported to reduce the incorporation of Sn.\cite{senaratne2014advances,bhargava2017doping,margetis2017fundamentals} Subsequently, detailed statistical analyses were performed on the APT dataset to probe the atomic behavior of Sn in the metastable PIN layers. The nearest-neighbor (NN) distribution was evaluated up to the $10^{\text{th}}$ order and compared with an ideal solution for the Sn-Sn pair of atoms. The analysis carried out on n-Ge$_{0.885}$Sn$_{0.115}$, i-Ge$_{0.90}$Sn$_{0.10}$, and p-Ge$_{0.92}$Sn$_{0.08}$ layers are shown in Fig. 1(f). The agreement between experimental (solid spheres) and APT simulations (dashed black curves) confirms the absence of Sn clusters in all GeSn layers, regardless of the nature of the dopants, as expected for a perfectly random alloy. These findings were further confirmed by the frequency distribution analysis (FDA) and partial radial distribution function (p-RDF) analysis. 

\begin{figure*}[t]
    \centering
    \includegraphics[scale=0.77]{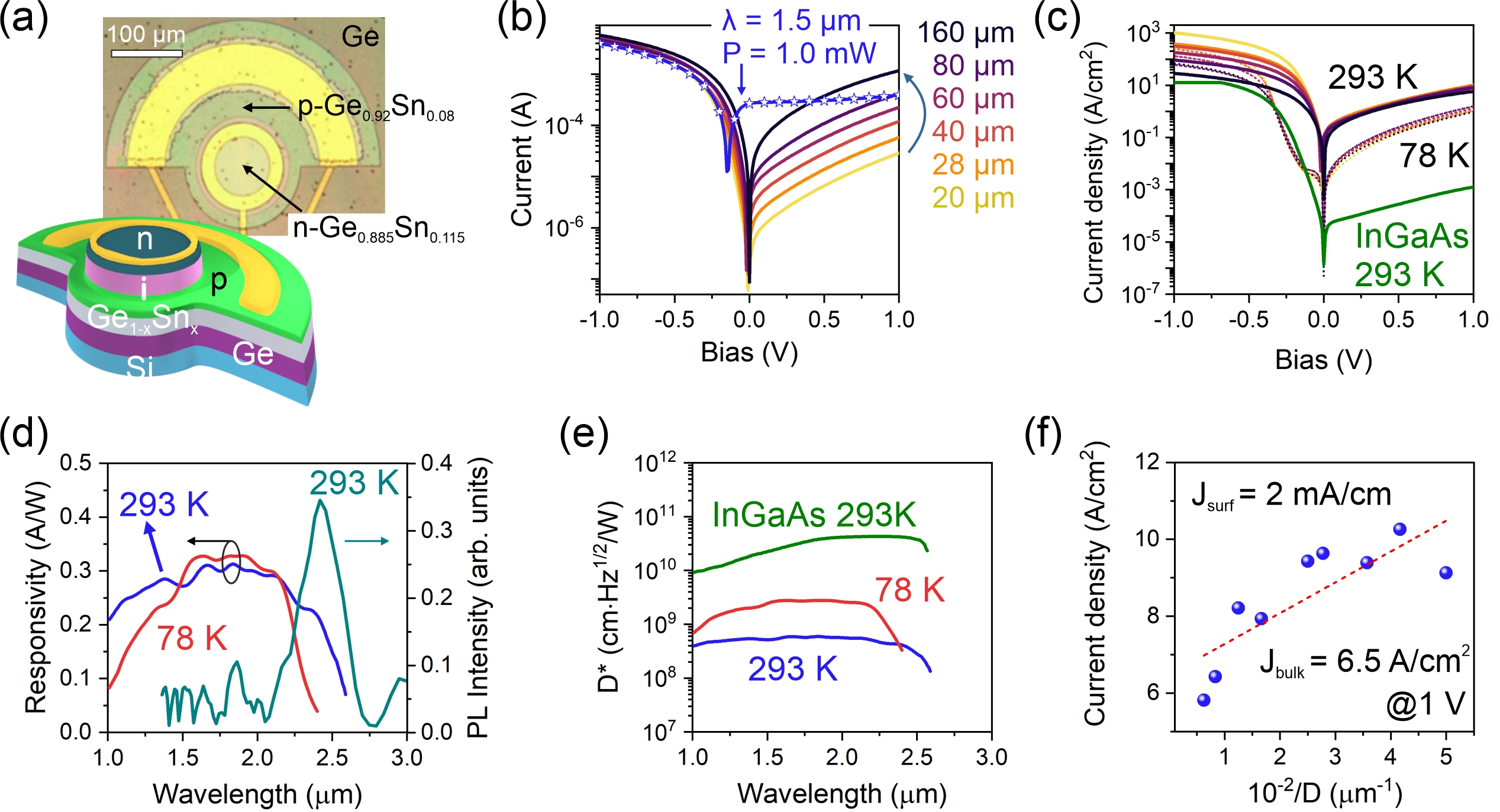}
    \caption{{\bf GeSn PIN photodetectors.} (a) Schematic and optical micrograph of a representative PIN PD. (b) I-V for dark current for various devices diameters along with I-V under illumination condition (blue line with symbols for $60$ $\mu$m device diameter). (c) I-V for dark current density at different device diameters at $293$ K and $78$ K compared to InGaAs at $293$ K. (d) Spectral responsivity of GeSn PIN device at RT and $78$ K along with RT PL for the as-grown sample. (e) Spectral detectivity of GeSn PIN device at $293$ K and $78$ K compared to that of InGaAs at $293$ K. (e) Dark current density as function of inverse of the device diameter at a reverse bias of $1$ V. The surface and bulk current densities can be extracted from the linear fit.}
\end{figure*}

\medskip

\noindent {\bf GeSn PIN photodetectors}

Vertical top-down PIN PDs with various mesa diameters were fabricated, as shown in Fig. 2(a). Etching using chlorine ICP to a depth of $0.914\,\mu$m was used to isolate each device from its neighboring ones and reduce the contact pads parasitic capacitance. A second chlorine etch is then made down to the p-layer with a depth of $450$ nm in the shape of circular bumps constituting the active area of the PD device. The devices have a varying diameter in the $20\text{--}160\,\mu$m range. To passivate the etched sidewall, chemical treatment in HCl:HF (1:1) solution for $10$ seconds was used followed by the deposition of SiO$_2$ layer. Openings are made in the SiO$_2$ layer for p- and n-contacts using BOE wet etch and then Ti/Au contacts were formed by e-beam evaporation. To discuss the device performance, it is important to first examine the I-V curves under dark and illumination conditions. To this end, the total current under $1.55\,\mu$m illumination at a power density of $50.9$ W/cm$^2$ is plotted as function of bias in Fig. 2(b). This corresponds to a room temperature responsivity of $0.3$ A/W at $0.5$ V for a $60\,\mu$m-diameter device. Similar responsivities were obtained for other devices at different diameters. It is noted that the minimum of the I-V curve under illumination occurs at a forward bias of $0.15$ V and not at zero bias owing to the built-in potential of the PIN junction. Dark current measurements show a rectification ratio higher than $10^2$ obtained at $0.5$ V for the $20\,\mu$m-diameter device (Fig. 2(b)), which indicates the relatively high quality of the GeSn PIN layers. The reverse dark current increases monotonically with the diameter, which is expected because of the increased importance of growth defects in larger devices. Note, however, that the reverse dark current densities measured at a variable bias are practically the same regardless of the device diameter (Fig. 2(c)), which is attributed to the low surface current leakage as the mesa diameter reduces from $160\,\mu$m to $20\,\mu$m. The dark current drops by an order of magnitude at a bias below $0.25$ V as the temperature is reduced to $78$ K, but it remains higher than that of a commercial InGaAs at room temperature. It is worth mentioning that the high responsivity at low dark current and low bias should significantly increase the detectivity of these devices as compared to photoconductive devices that suffer from high dark current and high noise.\cite{atalla2021all,tran2019si} This would allow the operation of GeSn PIN devices without the need for lock-in technique to extract the photocurrent signal.

\medskip

FTIR spectrometer was used to measure the relative spectral responsivity of the GeSn devices. The data was subsequently calibrated using an InGaAs detector with known spectral responsivity and a cutoff of $2.6\,\mu$m. As shown in Fig. 2(d), the responsivity of our devices reaches a value of $0.31$ A/W at $1.7\,\mu$m and plateaus up to $2.25\,\mu$m before slightly decreases as the wavelength increases with a rapid drop at $2.4\,\mu$m to reach a cutoff at $2.6\,\mu$m. This wavelength is consistent with the measured room-temperature photoluminescence (PL) of the as-grown heterostructure (Fig. 2(d)). The PL signal is broad with a peak centered at $2.4\,\mu$m, and its decay extends beyond $2.5\,\mu$m, which confirms the cutoff wavelength obtained from the responsivity measurements. 

\medskip

Another key characteristic of a PD is the specific detectivity D$^*$, which does not only account for the spectral responsivity but also for the dark current noise and the area of the device. D$^*$ is given by: D$^*\,=\sqrt{\text{A}}/$NEP,\cite{sze2021physics} where A is the PD  area and NEP is the noise equivalent power. To benchmark the performance of GeSn PDs, the obtained D$^*$ is compared to a commercial InGaAs detector (FD$10$D Thorlabs), as shown in Fig. 2(e). Considering the linear responsivity of these PDs and the absence of frequncey dependent noise, the NEP is calculated using:

\begin{equation}\label{NEPeq}
    \text{NEP}= I_{rms}/(\text{R$_\lambda$}\sqrt{\Delta f}),
\end{equation}

\noindent where R$_\lambda$ is the responsivity and $\Delta f$ is the frequency bandwidth and 
 
\begin{equation}\label{Irmseq}
    I_{rms}= \sqrt{(I_{therm}^2+ I_{shot}^2 )},
\end{equation}
 
\noindent where $I_{therm}$ and $I_{shot}$  are the thermal and shot noise currents, respectively.\cite{dong2017two} The thermal noise current is obtained using: $I_{therm} = \sqrt{(4\,kT\,\Delta f/R_{shunt})}$, where the shunt resistance $R_{shunt}$  is determined as the first derivative of the bias to dark current near $0$ V. For the $160\,\mu$m-diameter GeSn device, the $R_{shunt}=9.9$ k$\Omega$ and $I_{therm}/\sqrt{\Delta f}=1.28\times 10^{-12}$  A$\cdot$Hz$^{-1/2}$. The shot noise current, $I_{shot} = \sqrt{(2q\, (I_{dark}+I_{ph})\,\Delta f)}$, was calculated at $0.5$ V and R$_\lambda$ $= 0.3$ A/W as $9.88 \times10^{-12}$  A$\cdot$Hz$^{-1/2}$ for $1.55\,\mu$m illumination. This results in NEP $= 3.32\times10^{-11}$  W$\cdot$Hz$^{-1/2}$ and D$^* = 3.8\times10^8$  cm$\cdot$W$^{-1}\cdot$Hz$^{1/2}$ and $2.4\times10^9$  cm$\cdot$W$^{-1}\cdot$Hz$^{1/2}$ at $293$ K and $78$ K, respectively. For InGaAs detector at $293$ K, we found D$^* = 5\times10^{10}$  cm$\cdot$W$^{-1}\cdot$Hz$^{1/2}$. The spectral D$^*$ at $293$ K and $78$ K are plotted in Fig. 2(e) along with that of InGaAs. The main reason underlying the higher D$^*$ of InGaAs comes from its low dark current density as shown in Fig. 2(c).  This is not surprising as this technology has been benefiting from decades-long research on material and device optimization.\cite{dominici2016numerical} For GeSn, the current density at $78$ K shows a significant increase as the reverse bias increases indicating an electric-field enhanced trap-assisted thermionic emission owing to the induced band bending which depicts a Poole-Frenkel leakage effect.\cite{dong2017two,frenkel1938pre} It is noted that there are more than two orders of magnitude reduction in dark current density at $-\, 0.15$ V at $78$ K relative to that at $293$ K. This could be explained by the significant reduction in the Poole-Frenkel effect at a bias close to the built-in junction potential and the negligible phonon-assisted trap leakage. To separate the contributions of bulk leakage from surface leakage, the dark current density $J_{dark}$  as a function of inverse of device diameter D is plotted in Fig. 2(f) and fitted using the following linear equation: 

 \begin{equation}\label{Jdeq}
    J_{dark}= J_{bulk}+4\,J_{surf}/\text{D},
\end{equation}

\noindent where $J_{surf}$ is the surface leakage current density.\cite{zhou2020high} The dashed line shows the linear fit at a reverse bias of $1$ V. The obtained value of $J_{bulk}$ is $6.5$ A$/$cm$^2$ while $J_{surf}$ is just 2 mA$/$cm, which indicates the effective passivation of these PIN PDs. Consequently, only bulk leakage current should be considered in the analysis of the underlying mechanisms of dark current in these GeSn PDs. 

\begin{figure*}[t]
    \centering
    \includegraphics[scale=0.68]{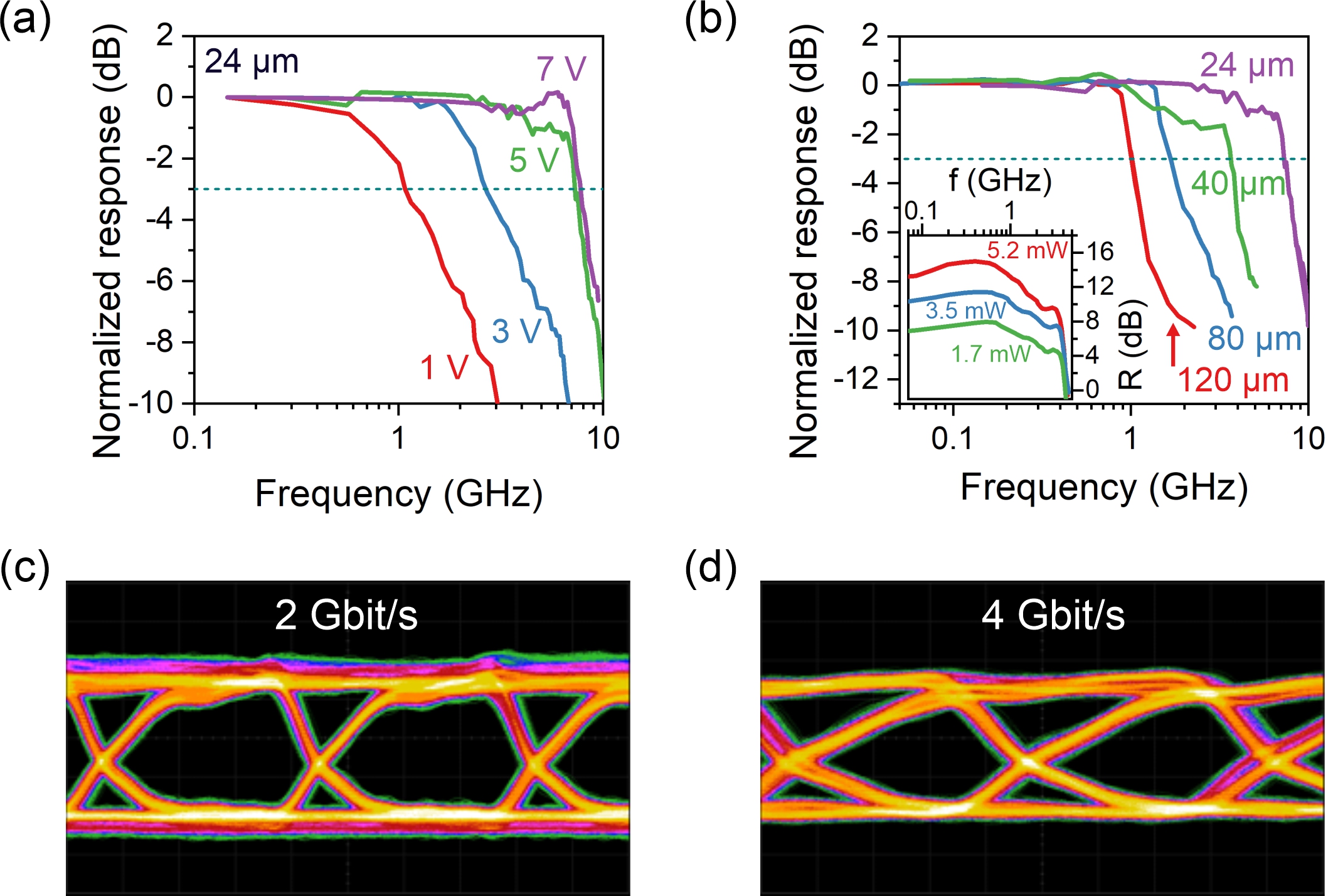}
    \caption{{\bf Photoresponse bandwidth of GeSn PDs.} (a) normalized photoresponse as function of optical pulse frequency depicting the bandwidth of the $24$ µm diameter device at various applied bias, (b) same as (a) but for various diameter devices at $5$ V reverse bias, (c) and (d) are the eye diagrams from the $24\,\mu$m device at $2$ Gbit/s and $4$ Gbit/s, respectively.}
\end{figure*}

\medskip

\noindent {\bf Photoresponse bandwidth of GeSn PDs}

The bandwidth is mainly limited by the carrier transit lifetime through the i-layer, the resistance capacitance (RC) delay, the diffusion current of the photocarriers generated in the n- and p-layers, and the carrier trapping at the heterojunction interfaces.\cite{tran2019study,zhou2020high} The last two factors would reduce the bandwidth of the device; however, they can be ignored because the photogenerated carriers in the n- and p-layers tend to recombine before reaching the contacts given the short carriers lifetime, and as the reverse bias increases the carrier trapping significantly reduces. The $3$\,dB bandwidth, $f_{-3\,\text{dB}}$, is given by:\cite{dong2017two,lin2017high}

\begin{equation}\label{freqeq}
    f_{-3\,dB}=\frac{1}{\sqrt{f_{Tr}^{-2}+f_{RC}^{-2}}},
\end{equation}

\noindent where $f_{Tr}=0.45\,\nu_s/t_i$ is the transit lifetime-limited bandwidth with $\nu_s$ and $t_i$ are the saturation velocity and thickness of the i-layer, respectively. $f_{RC}=1/(2\pi\text{RC})$ is the RC-limited bandwidth with R is the total resistance including series and load resistances, and C is the total capacitance including junction and parasitic capacitances. Assuming $\nu_s$ of GeSn equal to that of Ge and with $t_i=300$ nm, the $f_{Tr}$  will limit the PD bandwidth to $90$ GHz. For the $f_{RC}$, the series resistance was determined from the forward bias I-V curve to be $61.3\,\Omega$ and the load resistance was $50\,\Omega$. The device capacitance was measured for a $24\,\mu$m-diameter device to be $0.18\,$pF at $5$\,V yielding $f_{RC}=7.9$ GHz.

\medskip

Fig. 3(a) shows the normalized response measured using a $1.55\,\mu$m laser for the $24\,\mu$m device at various bias values with a dashed line marking the $-3\,$dB value of the response (see supplementary information for bandwidth measurement setup). It is noteworthy that $f_{RC}$  and $f_{Tr}$  are independent of wavelength, and these devices shall exhibit similarly high bandwidth above $2\,\mu$m wavelength since the cutoff is at $2.6\,\mu$m.\cite{zhou2020high} It is noticeable that, as the bias increases, the $-3$\,dB value increases to $1.08$ GHz, $2.7$ GHz, $7.5$ GHz, and $7.7$ GHz, at $1$ V, $3$ V, $5$ V, and $7$ V, respectively. The narrower bandwidth at low bias can be attributed to the high capacitance and carrier trapping at the heterojunction interfaces. As the bias increases, the device bandwidth increases but remains below the estimated value most likely because of the slow diffusion current component contributing to the total device photoresponse. In Fig. 3(b), the normalized bandwidths at various device diameters are compared at $5$ V. For the $120\,\mu$m, $80\,\mu$m, $40\,\mu$m and $24\,\mu$m devices, the measured $-3$\,dB bandwidth at $5$ V is $1$ GHz, $1.7$ GHz, $3.7$ GHz, and $7.5$ GHz, respectively. Note that the larger the device diameter, the larger the device capacitance, the narrower the bandwidth becomes. The inset in Fig. 3(b) shows the modulation of the photoresponse curve with the incident optical power at $5$ V indicating the bandwidth of the device is preserved as the illumination power changes. 

\medskip

In digital applications, the high bandwidth is usually demonstrated by an open eye diagram at various input signal frequencies because they can measure the additive noise in the RF signal, the dispersion and jitter and inter-bit interference. In Fig. 3(c) and 3(d), the open eye diagrams were measured for the highest bandwidth device at data rates of $2$ Gbit/s and $4$ Gbit/s. Note that the fact that the diagram is less clearly open at the $4$ Gbit/s is a limitation of the bandwidth of the pseudo-random bit generator and not related to the intrinsic performance of GeSn PDs. The open eye diagrams indicate low inter-bit interference and low signal distortion. 

\medskip

\begin{figure*}[t]
    \centering
    \includegraphics[scale=0.77]{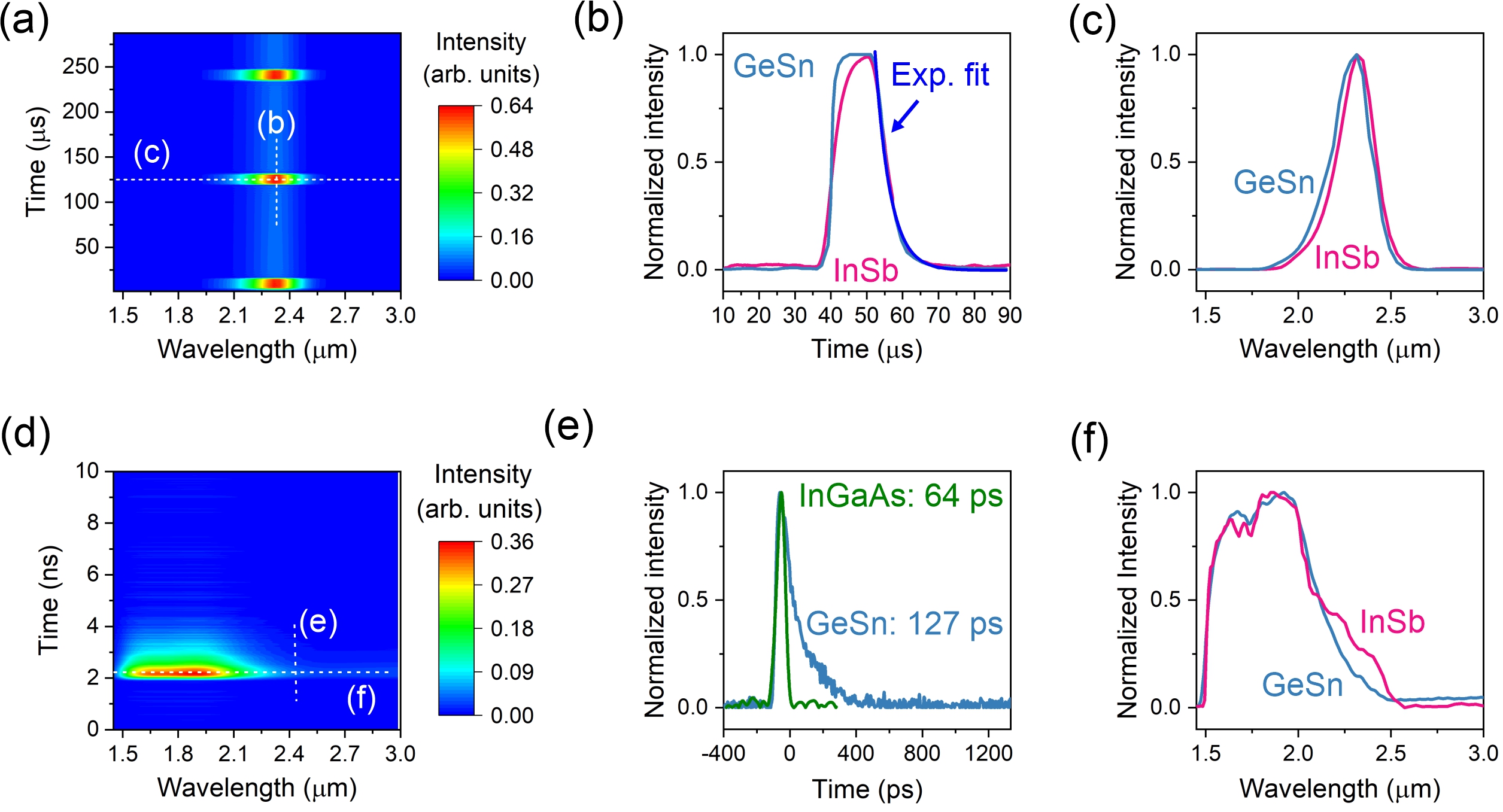}
    \caption{{\bf Time-resolved spectroscopy.} (a) 2D density plot of the spectral emission of a $2.35\,\mu$m LED and its temporal response under pulse drive current injection. (b) The waveform of the LED emission as measured by GeSn PD compared to liquid nitrogen (LN) cooled InSb PD. (c) Comparison between the LED emission spectrum measured by GeSn and InSb PDs and it shows the EL emission peak at $2.35\,\mu$m. (d) 2D density plot of the spectral emission of an ultra-short supercontinuum laser and its temporal response. (e) The waveform of the emission as measured by the GeSn PD compared to a high bandwidth InGaAs PD. (f) Comparison between spectrum measured by GeSn and LN cooled InSb PDs indicating the cutoff at $2.5\,\mu$m.}
\end{figure*}

\medskip

\noindent {\bf Integration of GeSn PIN PDs in ultrafast spectroscopy schemes}

The device characteristics outlined above hint to the potential of GeSn PDs for silicon-compatible e-SWIR applications. Yet, it is important to note that the high dark current (Fig. 2(c)) constitutes a major hurdle that needs to be overcome. This limitation calls for further improvement of the material quality.\cite{moutanabbir2021mono} Nonetheless, the demonstrated high-speed operation can still be relevant for applications that are not necessarily affected by the relatively high dark current. This includes time-resolved spectroscopy, where the typical excitation reaching the detector can be sufficiently intense. With this perspective, as a first demonstration, the GeSn PIN PD was used as a receiver in a time-resolved electro-luminescence (TREL) setup to probe the emission from a $2.35\,\mu$m commercial LED (LED Microsensor NT).\cite{Technology} TREL spectroscopy does not only provide spectral information but also provide insights into the temporal decay of the LED emission when a pulse driving current is applied to the LED. This technique can provide valuable information on the properties of the optoelectronic emitting devices.\cite{terada2013time,shatalov2003time,lin2011transient} In this demonstration, an LED driver is used to supply current pulse to the LED with $16\,\mu$s ON and $110\,\mu$s OFF. The LED emission passes through the FTIR spectrometer and redirected to fall on the GeSn PD afterwards. The GeSn PD is connected to the electronics of the spectrometer to generate time-resolved interferogram that is transformed into time-resolved spectrum, as depicted in Fig. 4(a). Comparing the spectra when current pulse is ON to that when it is OFF, ideally, the LED emission during the pulse OFF shall be zero. This LED is made of a double heterostructure comprising, in a top-down sequence, p-AlGaAsSb, n-GaInAsSb  (electron well) grown on n-GaSb substrate.\cite{Technology,danilova2005light} The LED emission at current OFF is likely because of the slow carrier diffusion into the heterostructure well at high current injection levels. A comparison with the results of an amplified liquid nitrogen (LN) cooled InSb PD is presented temporal-wise and spectral-wise in Figs. 4(b) and 4(c), respectively. In Fig 4(b), the InSb PD limited the rise time to its bandwidth, however GeSn shows much shorter rise time that is limited by the RC delay of the LED. On the contrary, a good match was found between LN cooled InSb and uncooled GeSn PDs in the fall time of the LED and it can be conveniently fitted to exponential decay function

\begin{equation}\label{yteq}
    y(t)=y_0+B\,e^{-\frac{(t- t_0)}{t_1}},
\end{equation}

\noindent where $y_0$ and $B$ are constants, $t_0=52.5\,\mu$s is the start of the fitting curve and $t_1=4.1\,\mu$s is the extracted LED carrier lifetime. The spectral emission intensity of the LED is presented in Fig. 4(c), and it confirms EL peak at $2.35\,\mu$m, measured by both LN cooled InSb and GeSn PDs and the slight difference in spectra is most likely caused by the difference in the spectral responsivity of the two PDs. 

\medskip

The e-SWIR high-speed PDs are also much needed for time-resolved spectroscopy (TRS)  to investigate transient biophysical and chemical processes.\cite{weinacht2018time,ritter2015time,smith2002fast} TRS techniques include FTIR interferometric spectroscopy and non-interferometric dispersive and laser-based approaches.\cite{ritter2015time} Although the laser-based methods have achieved extremely high time resolution,\cite{lee2008characterization} FTIR step-scan TRS features several advantages including ultra-wide spectral range, high resolution and high sensitivity, modest equipment cost and ease of use.\cite{kuhne2015early,boden2021investigation} Although interferometric techniques can measure the whole spectrum at once, they are limited to the speed of the scanning mirror (rapid scan) or the rise time of the photodetector (step scan).\cite{smith2002fast}  Since IR detectors with a cutoff above $2.3\,\mu$m lack the high bandwidth operation, the interferometeric techniques are limited to the nanosecond time resolution range.\cite{ritter2015time,smith2002fast} To tackle this limitation, GeSn PDs were integrated in an FTIR step-scan measurement with ps ultra-short pulse supercontinuum laser as a light source operating at wavelengths up to $2.5\,\mu$m. Fig. 4(d) shows a map of the photoresponse intensity  for the supercontinuum laser as a function of time and wavelength. In Fig. 4(e), the pulse waveform is measured with a $12$ GHz InGaAs ($1.7\,\mu$m cutoff) and shows $64$ ps FWHM, whereas a $127$ ps FWHM was obtained by GeSn PDs, which matched the $7.5$ GHz bandwidth measured for these devices. The cutoff of the laser is $2.5\,\mu$m as measured by both LN cooled InSb and GeSn detectors (in Fig. 4(f)). It is important to note that the obtained high time resolution using interferometric technique was heretofore not viable at wavelengths above $2.3\,\mu$m because of the lack of high bandwidth PD working in this range.

\medskip

\noindent {\bf Conclusion}

High frequency, silicon-integrated $2.6\,\mu$m detectors were achieved by exploiting the flexibility in bandgap engineering enabled by group IV GeSn semiconductors. All-GeSn PIN heterostructures were monolithically grown on silicon wafers using a step-graded growth process to tailor strain and content in the device layers and cover the e-SWIR range. Atomic-level studies revealed the high crystalline quality of the device stacks and confirmed the absence of any Sn clustering despite the far-from-equilibrium content. The obtained free-space detectors allow room temperature operation with a high responsivity of $0.3$ A/W and a bandwidth of $7.5$ GHz. Additionally, these characteristics were exploited in ultrafast spectroscopy to achieve time-resolved electro-luminescence and extract the carrier lifetime of a $2.3\,\mu$m light emitting diode. The high-speed detectors were also employed to reveal pulse duration, intensity, and spectral distribution of a pulsed supercontinuum laser using an FTIR spectrometer reaching a temporal resolution in the picosecond range at $2.5\,\mu$m. These demonstrated capabilities lay the groundwork to implement the long-sought-after scalable, cost-effective, and CMOS-compatible e-SWIR technologies.

\bigskip
\noindent {\bf Methods}

\noindent {\bf GeSn PIN layers epitaxial growth.} GeSn layers were grown on a 4-inch Si (100) wafer in chemical vapor deposition (CVD) reactor using ultra-pure H$_2$ carrier gas. $10\%$ monogermane (GeH$_4$) and tin-tetrachloride (SnCl$_4$) precursors, while p-type and n-type doping were controlled using $1.6\%$ Diborane (B$_2$H$_6$) and $1$\% Arsine (AsH$_3$). A $1.9\,\mu$m-thick Ge-VS was initially grown following a two-temperature growth process ($450$/$600$ $\,^\circ\text{C}$) and a post-growth thermal cyclic annealing ($>$ $800\,^\circ\text{C}$). The GeSn layers were grown at a reactor pressure of $50$ Torr, a constant H$_2$ flow, and a GeH$_4$ molar fraction ($1.2\times10^{-2}$). The layer composition was controlled by the temperature change.\cite{assali2018atomically,assali2019enhanced} In addition, the initial molar fraction of SnCl$_4$ ($9.1\times10^{-6}$) was reduced by $\sim20$\% during each temperature step to compensate for the reduced GeH$_4$ decomposition as temperature decreases. Four GeSn buffer layers with compositions of $5$ at.\%, $6$ at.\%, $8$ at.\%, and $9$ at.\% were grown at $335\,^\circ\text{C}$ (\#$1$), $325\,^\circ\text{C}$ (\#$2$), $315\,^\circ\text{C}$ (\#$3$) and $305\,^\circ\text{C}$ (\#$4$), respectively. The subsequent GeSn PIN growth ($8\text{--}9$\,/\,$10\text{--}11.5$ at.\%) was performed at a constant temperature of $305\,^\circ\text{C}$ and a fixed GeH$_4$ and SnCl$_4$ precursors supply (Ge/Sn ratio in gas phase $\sim2250$). B$_2$H$_6$ or AsH$_3$ were introduced for the growth of the p-GeSn and n-GeSn regions, respectively. A B$_2$H$_6$ supply with a B/Ge ratio in the gas phase of $\sim9\times10^{-5}$ was selected leading to an active doping significantly above $1\times10^{19}$ cm$^{-3}$. The same order of doping level was obtained using AsH$_3$ a supply with an As/Ge ratio in the gas phase of $\sim3\times10^{-3}$. The interfacial width (w) in the heterostructure are evaluated by fitting the compositional profiles with a sigmoidal function:\cite{dyck2017accurate}

\begin{equation}\label{zxeq}
    z(x)=K+ \frac{N}{1 + e^{-\left(x_0\pm x\right)/\tau}},
\end{equation}

\noindent where $K$ is a vertical offset parameter (related to the atomic content in the layer), $N$ is a scaling parameter (maximum atomic content), $x_0$ is the inflection point of the curve, and the sign of $x$ results in an increasing or a decreasing function. The interface width w is then estimated as $4\tau$.

\noindent {\bf Dark and photocurrent measurements.} The I-V measurements were acquired using Keithley 4200a parameter analyzer connected to a probe station. The photocurrent was measured at $1.55\,\mu$m wavelength showing a strong rectifying behavior especially for the small diameter devices. Moreover, the spectral responsivity was measured using Bruker Vertex 80 FTIR spectrometer. The IR light source of the spectrometer was incident on the GeSn device, and the electrical signal was measured using a Zurich Instruments lock-in amplifier that was locked to the frequency of a chopper in the light path of the light source. The lock-in signal is fed to the spectrometer electronics to eventually get the photocurrent as function of wavelength. Knowing the power profile of the light source the spectral responsivity of the PD can be calculated. The low temperature I-V measurements were doing using a Janis cryo-probe station and the light was injected onto the device via optical fiber and a focusing parabolic mirror.

\noindent {\bf Device bandwidth measurements.} Polarization maintained single frequency $1.55\,\mu$m laser was used and it was fiber coupled to a high bandwidth optical modulator connected to the Janis cryo-probe station and a focuser lens inside the probe station was used to incident light on the GeSn PDs. The electrical signal was collected using a microwave probe connected to a bias Tee that allow to DC bias the PD while extracting the AC photocurrent signal. That signal is then amplified and connected to a high bandwidth network analyzer while supplying the RF electric modulation signal to the optical modulator.

\noindent {\bf Time-resolved spectroscopy.} For picosecond temporal resolution, commercial FTIR spectrometers cannot be used primarily to perform TRS. Instead, the GeSn PD shall be connected to a tee bias, RF amplifier and external high bandwidth oscilloscope that was synchronized to record the waveforms as the FTIR mirror moves in steps inside the spectrometer. The collected waveforms form the oscilloscope will form a time-dependent interferogram that can be transformed into time-dependent spectrum using a Fourier transform Matlab code. A schematic depicting the TRS setup is provided in the supplementary information.

\bigskip
\noindent {\bf Acknowledgements}.
The authors thank Prof. S. Kena-Cohen for fruitful discussion and J. Bouchard for the technical support with the CVD system, O.M. acknowledges support from NSERC Canada (Discovery, SPG, and CRD Grants), Canada Research Chairs, Canada Foundation for Innovation, Mitacs, PRIMA Québec, and Defence Canada (Innovation for Defence Excellence and Security, IDEaS). 

\medskip

\noindent {\bf Author contributions}.
SA carried out the epitaxial growth. MRMA performed device fabrication and optoelectronic characterization including the bandwidth and TRS measurements. SK carried out TEM and APT analyses. AA calculated PIN device bandgap lineup using $k\cdot p$ simulations. OM led this research and helped in writing the manuscript with input from  MRMA, SA, SK and AA. All authors commented on the manuscript.

\medskip
\noindent {\bf Author information}.
Correspondence and requests for materials should be addressed to~:\\ oussama.moutanabbir@polymtl.ca

\smallskip
The authors declare no competing financial interest.

\medskip
\noindent {\bf Data availability}.
The data that support the findings of this study are available from the corresponding author upon reasonable request.

\bibliographystyle{naturemag}
\bibliography{main}

\begin{thebibliography}{10}
\expandafter\ifx\csname url\endcsname\relax
  \def\url#1{\texttt{#1}}\fi
\expandafter\ifx\csname urlprefix\endcsname\relax\def\urlprefix{URL }\fi
\providecommand{\bibinfo}[2]{#2}
\providecommand{\eprint}[2][]{\url{#2}}

\bibitem{wun2016}
\bibinfo{author}{Wun, J.-M.}, \bibinfo{author}{Wang, Y.-W.},
  \bibinfo{author}{Chen, Y.-H.}, \bibinfo{author}{Bowers, J.~E.} \&
  \bibinfo{author}{Shi, J.-W.}
\newblock \bibinfo{title}{\text{GaSb}-based pin photodiodes with partially
  depleted absorbers for high-speed and high-power performance at $2.5\,\mu$m
  wavelength}.
\newblock \emph{\bibinfo{journal}{IEEE Transactions on Electron Devices}}
  \textbf{\bibinfo{volume}{63}}, \bibinfo{pages}{2796--2801}
  (\bibinfo{year}{2016}).

\bibitem{Williams2017}
\bibinfo{author}{Williams, G.~M.}
\newblock \bibinfo{title}{{Optimization of eyesafe avalanche photodiode lidar
  for automobile safety and autonomous navigation systems}}.
\newblock \emph{\bibinfo{journal}{Optical Engineering}}
  \textbf{\bibinfo{volume}{56}}, \bibinfo{pages}{1 -- 9}
  (\bibinfo{year}{2017}).
\newblock \urlprefix\url{https://doi.org/10.1117/1.OE.56.3.031224}.

\bibitem{Kim2021}
\bibinfo{author}{Kim, I.} \emph{et~al.}
\newblock \bibinfo{title}{Nanophotonics for light detection and ranging
  technology}.
\newblock \emph{\bibinfo{journal}{Nature Nanotechnology}}
  \textbf{\bibinfo{volume}{16}}, \bibinfo{pages}{508 -- 524}
  (\bibinfo{year}{2021}).

\bibitem{Ye2015}
\bibinfo{author}{Ye, N.} \emph{et~al.}
\newblock \bibinfo{title}{\text{AlInGaAs} surface normal photodiode for
  $2\,\mu$m optical communication systems}.
\newblock In \emph{\bibinfo{booktitle}{2015 IEEE Photonics Conference (IPC)}},
  \bibinfo{pages}{456--459} (\bibinfo{year}{2015}).

\bibitem{joshi2008high}
\bibinfo{author}{Joshi, A.} \& \bibinfo{author}{Becker, D.}
\newblock \bibinfo{title}{High-speed low-noise pin \text{InGaAs} photoreceiver
  at $2\,\mu$m wavelength}.
\newblock \emph{\bibinfo{journal}{IEEE Photonics Technology Letters}}
  \textbf{\bibinfo{volume}{20}}, \bibinfo{pages}{551--553}
  (\bibinfo{year}{2008}).

\bibitem{yang2013}
\bibinfo{author}{Yang, H.} \emph{et~al.}
\newblock \bibinfo{title}{Butterfly packaged high-speed and low leakage
  \text{InGaAs} quantum well photodiode for $2000\,$nm wavelength systems}.
\newblock \emph{\bibinfo{journal}{Electronics letters}}
  \textbf{\bibinfo{volume}{49}}, \bibinfo{pages}{281--282}
  (\bibinfo{year}{2013}).

\bibitem{chen2018}
\bibinfo{author}{Chen, Y.} \emph{et~al.}
\newblock \bibinfo{title}{Dynamic model and bandwidth characterization of
  \text{InGaAs/GaAsSb} type-\text{II} quantum wells pin photodiodes}.
\newblock \emph{\bibinfo{journal}{Optics Express}}
  \textbf{\bibinfo{volume}{26}}, \bibinfo{pages}{35034--35045}
  (\bibinfo{year}{2018}).

\bibitem{tossoun2018}
\bibinfo{author}{Tossoun, B.} \emph{et~al.}
\newblock \bibinfo{title}{\text{InP}-based waveguide-integrated photodiodes
  with \text{InGaAs/GaAsSb} type-\text{II} quantum wells and 10-\text{GHz}
  bandwidth at $2\,\mu$m wavelength}.
\newblock \emph{\bibinfo{journal}{Journal of Lightwave Technology}}
  \textbf{\bibinfo{volume}{36}}, \bibinfo{pages}{4981--4987}
  (\bibinfo{year}{2018}).

\bibitem{chen2019hi}
\bibinfo{author}{Chen, Y.}, \bibinfo{author}{Xie, Z.}, \bibinfo{author}{Huang,
  J.}, \bibinfo{author}{Deng, Z.} \& \bibinfo{author}{Chen, B.}
\newblock \bibinfo{title}{High-speed uni-traveling carrier photodiode for 2
  $\mu$m wavelength application}.
\newblock \emph{\bibinfo{journal}{Optica}} \textbf{\bibinfo{volume}{6}},
  \bibinfo{pages}{884--889} (\bibinfo{year}{2019}).

\bibitem{andreev2013high}
\bibinfo{author}{Andreev, I.} \emph{et~al.}
\newblock \bibinfo{title}{High-speed photodiodes for the mid-infrared spectral
  region 1.2\text{--}2.4 $\mu$m based on \text{GaSb/GaInAsSb/GaAlAsSb}
  heterostructures with a transmission band of 2\text{--}5 \text{GHz}}.
\newblock \emph{\bibinfo{journal}{Semiconductors}}
  \textbf{\bibinfo{volume}{47}}, \bibinfo{pages}{1103--1109}
  (\bibinfo{year}{2013}).

\bibitem{geis2007cmos}
\bibinfo{author}{Geis, M.} \emph{et~al.}
\newblock \bibinfo{title}{\text{CMOS}-compatible \text{all-Si} high-speed
  waveguide photodiodes with high responsivity in near-infrared communication
  band}.
\newblock \emph{\bibinfo{journal}{IEEE Photonics Technology Letters}}
  \textbf{\bibinfo{volume}{19}}, \bibinfo{pages}{152--154}
  (\bibinfo{year}{2007}).

\bibitem{grote201210}
\bibinfo{author}{Grote, R.~R.} \emph{et~al.}
\newblock \bibinfo{title}{\text{10 Gb/s} error-free operation of all-silicon
  ion-implanted-waveguide photodiodes at $1.55\,\mu$m}.
\newblock \emph{\bibinfo{journal}{IEEE Photonics Technology Letters}}
  \textbf{\bibinfo{volume}{25}}, \bibinfo{pages}{67--70}
  (\bibinfo{year}{2012}).

\bibitem{souhan2014si+}
\bibinfo{author}{Souhan, B.} \emph{et~al.}
\newblock \bibinfo{title}{Si$^+$-implanted \text{Si-wire} waveguide
  photodetectors for the mid-infrared}.
\newblock \emph{\bibinfo{journal}{Optics Express}}
  \textbf{\bibinfo{volume}{22}}, \bibinfo{pages}{27415--27424}
  (\bibinfo{year}{2014}).

\bibitem{ackert2015high}
\bibinfo{author}{Ackert, J.~J.} \emph{et~al.}
\newblock \bibinfo{title}{High-speed detection at two micrometres with
  monolithic silicon photodiodes}.
\newblock \emph{\bibinfo{journal}{Nature Photonics}}
  \textbf{\bibinfo{volume}{9}}, \bibinfo{pages}{393--396}
  (\bibinfo{year}{2015}).

\bibitem{soref2010mid}
\bibinfo{author}{Soref, R.}
\newblock \bibinfo{title}{Mid-infrared photonics in silicon and germanium}.
\newblock \emph{\bibinfo{journal}{Nature Photonics}}
  \textbf{\bibinfo{volume}{4}}, \bibinfo{pages}{495--497}
  (\bibinfo{year}{2010}).

\bibitem{moutanabbir2021mono}
\bibinfo{author}{Moutanabbir, O.} \emph{et~al.}
\newblock \bibinfo{title}{Monolithic infrared silicon photonics: the rise of
  \text{(Si) GeSn} semiconductors}.
\newblock \emph{\bibinfo{journal}{Applied Physics Letters}}
  \textbf{\bibinfo{volume}{118}}, \bibinfo{pages}{110502}
  (\bibinfo{year}{2021}).

\bibitem{soref2015enabling}
\bibinfo{author}{Soref, R.}
\newblock \bibinfo{title}{Enabling 2 $\mu$m communications}.
\newblock \emph{\bibinfo{journal}{Nature Photonics}}
  \textbf{\bibinfo{volume}{9}}, \bibinfo{pages}{358--359}
  (\bibinfo{year}{2015}).

\bibitem{xu2019high}
\bibinfo{author}{Xu, S.} \emph{et~al.}
\newblock \bibinfo{title}{High-speed photodetection at two-micron-wavelength:
  technology enablement by \text{GeSn/Ge} multiple-quantum-well photodiode on
  300 mm \text{Si} substrate}.
\newblock \emph{\bibinfo{journal}{Optics Express}}
  \textbf{\bibinfo{volume}{27}}, \bibinfo{pages}{5798--5813}
  (\bibinfo{year}{2019}).

\bibitem{tran2019study}
\bibinfo{author}{Tran, H.} \emph{et~al.}
\newblock \bibinfo{title}{Study of \text{GeSn} mid-infrared photodetectors for
  high frequency applications}.
\newblock \emph{\bibinfo{journal}{Frontiers in Materials}}
  \textbf{\bibinfo{volume}{6}}, \bibinfo{pages}{278} (\bibinfo{year}{2019}).

\bibitem{talamas2021}
\bibinfo{author}{Talamas~Simola, E.} \emph{et~al.}
\newblock \bibinfo{title}{\text{CMOS}-compatible bias-tunable dual-band
  detector based on \text{GeSn/Ge/Si} coupled photodiodes}.
\newblock \emph{\bibinfo{journal}{ACS Photonics}} \textbf{\bibinfo{volume}{8}},
  \bibinfo{pages}{2166--2173} (\bibinfo{year}{2021}).

\bibitem{elbaz2020ultra}
\bibinfo{author}{Elbaz, A.} \emph{et~al.}
\newblock \bibinfo{title}{Ultra-low-threshold continuous-wave and pulsed lasing
  in tensile-strained \text{GeSn} alloys}.
\newblock \emph{\bibinfo{journal}{Nature Photonics}}
  \textbf{\bibinfo{volume}{14}}, \bibinfo{pages}{375--382}
  (\bibinfo{year}{2020}).

\bibitem{zhou2020elec}
\bibinfo{author}{Zhou, Y.} \emph{et~al.}
\newblock \bibinfo{title}{Electrically injected \text{GeSn} lasers on \text{Si}
  operating up to 100 \text{K}}.
\newblock \emph{\bibinfo{journal}{Optica}} \textbf{\bibinfo{volume}{7}},
  \bibinfo{pages}{924--928} (\bibinfo{year}{2020}).

\bibitem{chtien2019}
\bibinfo{author}{Chretien, J.} \emph{et~al.}
\newblock \bibinfo{title}{\text{GeSn} lasers covering a wide wavelength range
  thanks to uniaxial tensile strain}.
\newblock \emph{\bibinfo{journal}{ACS Photonics}} \textbf{\bibinfo{volume}{6}},
  \bibinfo{pages}{2462--2469} (\bibinfo{year}{2019}).

\bibitem{aubin2017}
\bibinfo{author}{Aubin, J.} \emph{et~al.}
\newblock \bibinfo{title}{Growth and structural properties of step-graded, high
  \text{Sn} content \text{GeSn} layers on \text{Ge}}.
\newblock \emph{\bibinfo{journal}{Semiconductor Science and Technology}}
  \textbf{\bibinfo{volume}{32}}, \bibinfo{pages}{094006}
  (\bibinfo{year}{2017}).

\bibitem{koelling2020}
\bibinfo{author}{Koelling, S.} \emph{et~al.}
\newblock \bibinfo{title}{Probing semiconductor heterostructures from the
  atomic to the micrometer scale}.
\newblock \emph{\bibinfo{journal}{ECS Transactions}}
  \textbf{\bibinfo{volume}{98}}, \bibinfo{pages}{447} (\bibinfo{year}{2020}).

\bibitem{senaratne2014advances}
\bibinfo{author}{Senaratne, C.}, \bibinfo{author}{Gallagher, J.},
  \bibinfo{author}{Aoki, T.}, \bibinfo{author}{Kouvetakis, J.} \&
  \bibinfo{author}{Menendez, J.}
\newblock \bibinfo{title}{Advances in light emission from group-\text{IV}
  alloys via lattice engineering and n-type doping based on custom-designed
  chemistries}.
\newblock \emph{\bibinfo{journal}{Chemistry of Materials}}
  \textbf{\bibinfo{volume}{26}}, \bibinfo{pages}{6033--6041}
  (\bibinfo{year}{2014}).

\bibitem{bhargava2017doping}
\bibinfo{author}{Bhargava, N.}, \bibinfo{author}{Margetis, J.} \&
  \bibinfo{author}{Tolle, J.}
\newblock \bibinfo{title}{As doping of \text{Si-Ge-Sn} epitaxial semiconductor
  materials on a commercial \text{CVD} reactor}.
\newblock \emph{\bibinfo{journal}{Semiconductor Science and Technology}}
  \textbf{\bibinfo{volume}{32}}, \bibinfo{pages}{094003}
  (\bibinfo{year}{2017}).

\bibitem{margetis2017fundamentals}
\bibinfo{author}{Margetis, J.} \emph{et~al.}
\newblock \bibinfo{title}{Fundamentals of \text{Ge$_{1-x}$Sn$_x$} and
  \text{Si$_y$Ge$_{1-x-y}$Sn$_x$ RPCVD} epitaxy}.
\newblock \emph{\bibinfo{journal}{Materials Science in Semiconductor
  Processing}} \textbf{\bibinfo{volume}{70}}, \bibinfo{pages}{38--43}
  (\bibinfo{year}{2017}).

\bibitem{atalla2021all}
\bibinfo{author}{Atalla, M.~R.} \emph{et~al.}
\newblock \bibinfo{title}{All-group \text{IV} transferable membrane
  mid-infrared photodetectors}.
\newblock \emph{\bibinfo{journal}{Advanced Functional Materials}}
  \textbf{\bibinfo{volume}{31}}, \bibinfo{pages}{2006329}
  (\bibinfo{year}{2021}).

\bibitem{tran2019si}
\bibinfo{author}{Tran, H.} \emph{et~al.}
\newblock \bibinfo{title}{Si-based \text{GeSn} photodetectors toward
  mid-infrared imaging applications}.
\newblock \emph{\bibinfo{journal}{ACS Photonics}} \textbf{\bibinfo{volume}{6}},
  \bibinfo{pages}{2807--2815} (\bibinfo{year}{2019}).

\bibitem{sze2021physics}
\bibinfo{author}{Sze, S.~M.}, \bibinfo{author}{Li, Y.} \& \bibinfo{author}{Ng,
  K.~K.}
\newblock \emph{\bibinfo{title}{Physics of semiconductor devices}}
  (\bibinfo{publisher}{John Wiley \& Sons}, \bibinfo{year}{2021}).

\bibitem{dong2017two}
\bibinfo{author}{Dong, Y.} \emph{et~al.}
\newblock \bibinfo{title}{Two-micron-wavelength germanium-tin photodiodes with
  low dark current and gigahertz bandwidth}.
\newblock \emph{\bibinfo{journal}{Optics Express}}
  \textbf{\bibinfo{volume}{25}}, \bibinfo{pages}{15818--15827}
  (\bibinfo{year}{2017}).

\bibitem{dominici2016numerical}
\bibinfo{author}{Dominici, S.}, \bibinfo{author}{Wen, H.},
  \bibinfo{author}{Bertazzi, F.}, \bibinfo{author}{Goano, M.} \&
  \bibinfo{author}{Bellotti, E.}
\newblock \bibinfo{title}{Numerical study on the optical and carrier
  recombination processes in \text{GeSn} alloy for \text{E-SWIR} and
  \text{MWIR} optoelectronic applications}.
\newblock \emph{\bibinfo{journal}{Optics Express}}
  \textbf{\bibinfo{volume}{24}}, \bibinfo{pages}{26363--26381}
  (\bibinfo{year}{2016}).

\bibitem{frenkel1938pre}
\bibinfo{author}{Frenkel, J.}
\newblock \bibinfo{title}{On pre-breakdown phenomena in insulators and
  electronic semi-conductors}.
\newblock \emph{\bibinfo{journal}{Physical Review}}
  \textbf{\bibinfo{volume}{54}}, \bibinfo{pages}{647} (\bibinfo{year}{1938}).

\bibitem{zhou2020high}
\bibinfo{author}{Zhou, H.} \emph{et~al.}
\newblock \bibinfo{title}{High-efficiency \text{GeSn/Ge} multiple-quantum-well
  photodetectors with photon-trapping microstructures operating at 2 $\mu$m}.
\newblock \emph{\bibinfo{journal}{Optics Express}}
  \textbf{\bibinfo{volume}{28}}, \bibinfo{pages}{10280--10293}
  (\bibinfo{year}{2020}).

\bibitem{lin2017high}
\bibinfo{author}{Lin, Y.} \emph{et~al.}
\newblock \bibinfo{title}{High-efficiency normal-incidence vertical pin
  photodetectors on a germanium-on-insulator platform}.
\newblock \emph{\bibinfo{journal}{Photonics Research}}
  \textbf{\bibinfo{volume}{5}}, \bibinfo{pages}{702--709}
  (\bibinfo{year}{2017}).

\bibitem{Technology}
\bibinfo{title}{Technology}.
\newblock \bibinfo{howpublished}{\url{http://lmsnt.com/about-us/technology/}}.
\newblock \bibinfo{note}{Accessed: 2021-10-20}.

\bibitem{terada2013time}
\bibinfo{author}{Terada, Y.}, \bibinfo{author}{Yasutake, Y.} \&
  \bibinfo{author}{Fukatsu, S.}
\newblock \bibinfo{title}{Time-resolved electroluminescence of bulk \text{Ge}
  at room temperature}.
\newblock \emph{\bibinfo{journal}{Applied Physics Letters}}
  \textbf{\bibinfo{volume}{102}}, \bibinfo{pages}{041102}
  (\bibinfo{year}{2013}).

\bibitem{shatalov2003time}
\bibinfo{author}{Shatalov, M.} \emph{et~al.}
\newblock \bibinfo{title}{Time-resolved electroluminescence of
  \text{AlGaN}-based light-emitting diodes with emission at $285\,$nm}.
\newblock \emph{\bibinfo{journal}{Applied physics letters}}
  \textbf{\bibinfo{volume}{82}}, \bibinfo{pages}{167--169}
  (\bibinfo{year}{2003}).

\bibitem{lin2011transient}
\bibinfo{author}{Lin, M.-T.}, \bibinfo{author}{Li, M.}, \bibinfo{author}{Chen,
  W.-H.}, \bibinfo{author}{Omary, M.~A.} \& \bibinfo{author}{Shepherd, N.~D.}
\newblock \bibinfo{title}{Transient electroluminescence determination of
  carrier mobility and charge trapping effects in heavily doped phosphorescent
  organic light-emitting diodes}.
\newblock \emph{\bibinfo{journal}{Solid-State Electronics}}
  \textbf{\bibinfo{volume}{56}}, \bibinfo{pages}{196--200}
  (\bibinfo{year}{2011}).

\bibitem{danilova2005light}
\bibinfo{author}{Danilova, T.}, \bibinfo{author}{Zhurtanov, B.},
  \bibinfo{author}{Imenkov, A.} \& \bibinfo{author}{Yakovlev, Y.~P.}
\newblock \bibinfo{title}{Light-emitting diodes based on \text{GaSb} alloys for
  the 1.6--4.4 $\mu$m mid-infrared spectral range}.
\newblock \emph{\bibinfo{journal}{Semiconductors}}
  \textbf{\bibinfo{volume}{39}}, \bibinfo{pages}{1235--1266}
  (\bibinfo{year}{2005}).

\bibitem{weinacht2018time}
\bibinfo{author}{Weinacht, T.~C.} \& \bibinfo{author}{Pearson, B.~J.}
\newblock \emph{\bibinfo{title}{Time-resolved spectroscopy: An experimental
  perspective}} (\bibinfo{publisher}{CRC Press}, \bibinfo{year}{2018}).

\bibitem{ritter2015time}
\bibinfo{author}{Ritter, E.} \emph{et~al.}
\newblock \bibinfo{title}{Time-resolved infrared spectroscopic techniques as
  applied to channelrhodopsin}.
\newblock \emph{\bibinfo{journal}{Frontiers in molecular biosciences}}
  \textbf{\bibinfo{volume}{2}}, \bibinfo{pages}{38} (\bibinfo{year}{2015}).

\bibitem{smith2002fast}
\bibinfo{author}{Smith, G.~D.} \& \bibinfo{author}{Palmer, R.~A.}
\newblock \bibinfo{title}{Fast time-resolved mid-infrared spectroscopy using an
  interferometer}.
\newblock \emph{\bibinfo{journal}{Handbook of vibrational spectroscopy}}
  \textbf{\bibinfo{volume}{1}}, \bibinfo{pages}{625--640}
  (\bibinfo{year}{2002}).

\bibitem{lee2008characterization}
\bibinfo{author}{Lee, K.~F.}, \bibinfo{author}{Kubarych, K.~J.},
  \bibinfo{author}{Bonvalet, A.} \& \bibinfo{author}{Joffre, M.}
\newblock \bibinfo{title}{Characterization of mid-infrared femtosecond pulses}.
\newblock \emph{\bibinfo{journal}{JOSA B}} \textbf{\bibinfo{volume}{25}},
  \bibinfo{pages}{A54--A62} (\bibinfo{year}{2008}).

\bibitem{kuhne2015early}
\bibinfo{author}{Kuhne, J.} \emph{et~al.}
\newblock \bibinfo{title}{Early formation of the ion-conducting pore in
  channelrhodopsin-2}.
\newblock \emph{\bibinfo{journal}{Angewandte Chemie International Edition}}
  \textbf{\bibinfo{volume}{54}}, \bibinfo{pages}{4953--4957}
  (\bibinfo{year}{2015}).

\bibitem{boden2021investigation}
\bibinfo{author}{Boden, P.} \emph{et~al.}
\newblock \bibinfo{title}{Investigation of luminescent triplet states in
  tetranuclear \text{CuI} complexes: Thermochromism and structural
  characterization}.
\newblock \emph{\bibinfo{journal}{Chemistry (Weinheim an der Bergstrasse,
  Germany)}} \textbf{\bibinfo{volume}{27}}, \bibinfo{pages}{5439}
  (\bibinfo{year}{2021}).

\bibitem{assali2018atomically}
\bibinfo{author}{Assali, S.}, \bibinfo{author}{Nicolas, J.},
  \bibinfo{author}{Mukherjee, S.}, \bibinfo{author}{Dijkstra, A.} \&
  \bibinfo{author}{Moutanabbir, O.}
\newblock \bibinfo{title}{Atomically uniform \text{Sn}-rich \text{GeSn}
  semiconductors with 3.0--3.5\,$\mu$m room-temperature optical emission}.
\newblock \emph{\bibinfo{journal}{Applied Physics Letters}}
  \textbf{\bibinfo{volume}{112}}, \bibinfo{pages}{251903}
  (\bibinfo{year}{2018}).

\bibitem{assali2019enhanced}
\bibinfo{author}{Assali, S.}, \bibinfo{author}{Nicolas, J.} \&
  \bibinfo{author}{Moutanabbir, O.}
\newblock \bibinfo{title}{Enhanced \text{Sn} incorporation in \text{GeSn}
  epitaxial semiconductors via strain relaxation}.
\newblock \emph{\bibinfo{journal}{Journal of Applied Physics}}
  \textbf{\bibinfo{volume}{125}}, \bibinfo{pages}{025304}
  (\bibinfo{year}{2019}).

\bibitem{dyck2017accurate}
\bibinfo{author}{Dyck, O.} \emph{et~al.}
\newblock \bibinfo{title}{Accurate quantification of \text{Si/SiGe} interface
  profiles via atom probe tomography}.
\newblock \emph{\bibinfo{journal}{Advanced Materials Interfaces}}
  \textbf{\bibinfo{volume}{4}}, \bibinfo{pages}{1700622}
  (\bibinfo{year}{2017}).

\end{thebibliography}


\begin{figure*}[t]
    \centering
    \includegraphics[scale=0.86]{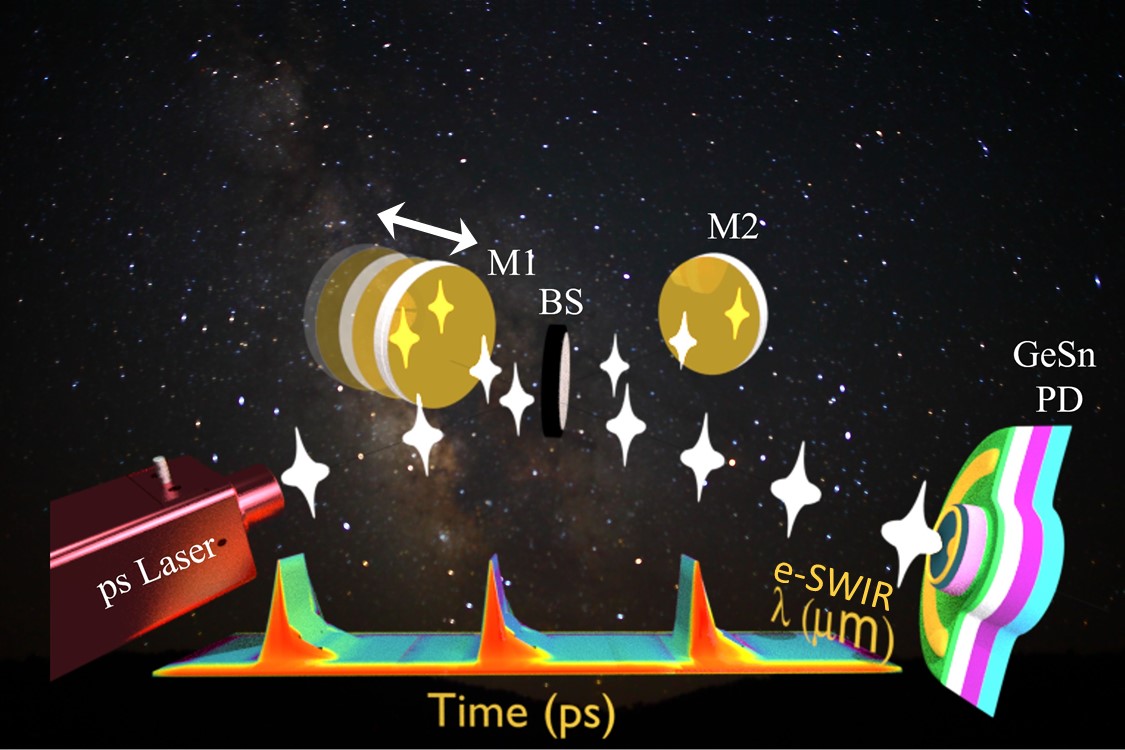}
    \caption{{\bf Table of content} }
\end{figure*}

\end{document}